\documentclass[12pt]{article}
\usepackage{amsmath}
\def\a{\alpha}

\def\d{\delta}
\def\e{\epsilon}

\def\m{\mu}
\def\n{\nu}

\def\be{\begin{equation}}
\def\ee{\end{equation}}
\def\arr{\begin{array}{rll}}
\def\ea{\end{array}}
\def\bea{\begin{eqnarray}}
\def\eea{\end{eqnarray}}

\def\N2{$N{=}2$}

\def\>{\rangle}
\def\<{\langle}
\def\+{\dagger}
\def\={\ =\ }

\begin{document}
\renewcommand{\thefootnote}{\fnsymbol{footnote}}
\begin{titlepage}
\setcounter{page}{0}
\vskip 1cm
\begin{center}
{\LARGE\bf Coset spaces and Einstein manifolds  }\\
\vskip 0.5cm
{\LARGE\bf with $l$--conformal Galilei symmetry }\\
\vskip 1cm
$
\textrm{\Large Dmitry Chernyavsky \ }
$
\vskip 0.7cm
{\it
Laboratory of Mathematical Physics, Tomsk Polytechnic University, \\
634050 Tomsk, Lenin Ave. 30, Russian Federation} \\
{E-mail: chernyavsky@tpu.ru}

\end{center}
\vskip 1cm
\begin{abstract} \noindent
The group theoretic construction is applied to construct a novel dynamical realization of the $l$--conformal Galilei group in terms of geodesic equations on the coset space.
A peculiar feature of the geodesics is that all their integrals of motion, including the accelerations, are functionally independent.
The analysis in the recent work [Phys. Lett. B 754 (2016) 249] is extended to construct the Einstein metrics with the $l$--conformal Galilei isometry group.
\end{abstract}


\vskip 0.5cm


\end{titlepage}

\renewcommand{\thefootnote}{\arabic{footnote}}
\setcounter{footnote}0

\noindent
{\bf 1. Introduction}\\

\noindent
During the last decade conformal extensions of the Galilei and Newton--Hooke algebras \cite{HP,{nor}} attracted considerable interest \cite{LSZ}--\cite{cg} primarily due to the current studies of the nonrelativistic version of AdS/CFT--correspondence. Galilei algebra admits a family of conformal extensions indexed by a positive integer or half-integer  $l$, known in the literature as the $l$--conformal Galilei algebra \cite{nor}. This algebra represents the semi direct sum of $so(2,1)\oplus so(d)$ and an abelian subalgebra generated by $d(2l+1)$ vector generators $C_i^{(n)}$ with $n=0,\dots,2l$ and $i=1,\dots,d$. The instances of $n=0,1$ correspond to spatial translations and Galilei boosts, while $n>1$ yield the so called acceleration generators.

One of the basic questions concerning any group is the construction of dynamical systems invariant under the action of the group. Dynamical realizations of the $l$-conformal Galilei/Newton--Hooke group have been extensively studied in \cite{LSZ}--\cite{cg}. When constructing a dynamical realization, integrals of motion are linked to the generators in the algebra. Because of the presence of the acceleration generators in the algebra,
most of the dynamical realizations known to date involve higher derivative terms. The second order systems for which the acceleration generators are functionally dependent have been constructed in \cite{GM3,GM1}. More recently,
a second order system for which all the generators in the $l$-conformal Galilei algebra are linked to functionally independent integrals of motion has been constructed \cite{cg} by considering geodesic equations on a Ricci--flat spacetime with the $l$-conformal Galilei isometry group. As compared to the previous studies, a novel feature is that to each acceleration generator in the algebra there correspond extra dimensions in spacetime. In other words, the dimension of spacetime in which the dynamical system is realized grows with $l$.

In this work we elaborate on the proposal in \cite{cg} and construct a similar dynamical realization by considering geodesic equations on a specially chosen coset space which need not be Ricci--flat. In contrast to \cite{cg}, the equations are given in explicit form. We also generalize the results in \cite{cg} to include the cosmological constant.

In the next section we construct the Maurer--Cartan one--forms on the coset space of the $l$--conformal Galilei group. These are used to build an invariant metric on the coset space. In Sect. 3 we analyze geodesic equations associated with such a metric in purely group--theoretic terms. Einstein manifolds which enjoy the $l$--conformal Galilei symmetry are discussed in Sect. 4. We summarize our results in the concluding Sect. 5.

\vspace{0.5cm}

\noindent
{\bf 2. Coset construction for the $l$-conformal Galilei algebra}\\

\noindent
The $l$--conformal Galilei algebra involves the generators of time translations $H$, dilatation $D$, special conformal transformation $K$, vector generators $C_i^{(n)}$ and spatial rotations $M_{ij}$. The structure relations read \cite{nor}
\bea
&&
[H,D]=iH, \qquad [H,K]=2i D, \qquad [D,K]=iK,
\nonumber\\[6pt]
&&
[H,C^{(n)}_i]=in C^{(n-1)}_i, \qquad\quad\;\; [D,C^{(n)}_i]=i(n-l) C^{(n)}_i,
\nonumber
\eea
\bea\label{algebra*}
&&
[K,C^{(n)}_i]=i(n-2l) C^{(n+1)}_i, \;\; [M_{ij},C^{(n)}_k]=-i\delta_{ik} C^{(n)}_j+i\delta_{jk} C^{(n)}_i,
\nonumber\\[6pt]
&&
[M_{ij},M_{kl}]=-i\delta_{ik} M_{jl}-i\delta_{jl} M_{ik}+
i\delta_{il} M_{jk}+i\delta_{jk} M_{il}.
\eea
The metric on the coset space $G/H$, where $G$ is the $l$--conformal Galilei group and $H$ is the subgroup generated by $D$ and $M_{ij}$, depends on the coordinates $t$, $r$ and $x^{(n)}_i$ with $x^{(-1)}_i=x^{(2l+1)}_i=0$ and is constructed by analogy with Ref. \cite{cg}:
\be\label{metr}
ds^2= \omega_H\omega_K+S_{n,m} \omega^{(n)}_i \omega^{(m)}_i,
\ee
where we denoted
\bea\label{MC}
&&
\omega_H=dt, \qquad\omega_K=r^2 dt+dr, \qquad \omega_D=-2r dt, \qquad \omega^{(n)}_i=d x^{(n)}_i+a^{(n)}_i dt+b^{(n)}_i dr,
\\[4pt]
&&
a^{(n)}_i=2r (n-l)x^{(n)}_i-(n+1)x^{(n+1)}_i-r^2(n-2l-1)x^{(n-1)}_i, \quad b^{(n)}_i=-(n-2l-1)x^{(n-1)}_i.
\nonumber
\eea
The constant matrix $S_{m,n}$ is off--diagonal and obeys the condition
\be\label{scond}
S_{m,n}(m+n-2l)=0, \quad \forall ~m,~n,
\ee
providing the invariance of the metric under the action of the $l$--conformal Galilei group. Note that for the case of $d=1$, $l=1$ conformal Galilei algebra
similar metrics have been studied in \cite{BK}.

\vspace{0.5cm}

\noindent
{\bf 2. Geodesics equations}\\

\noindent
The geodesic equations involve the Christoffel symbols associated, which may have rather bulky form. Yet,
for geodesics on coset spaces there exists a natural way to treat them in group-theoretic terms.

Consider a $(p+q)$--dimensional Lie group $G$ with a $q$--dimensional subgroup $H$. Let the corresponding Lie algebra be spanned by the generators
 $P_a (a=1,...,p)$ and  $M_i (i=1,...,q)$ with the structure relations
\bea\label{alg}
&&
[M_i,M_j]=f_{ij}^kM_k, \quad [P_a,P_b]=f_{ab}^kM_k+f_{ab}^c P_c, \quad [M_i,P_a]=f_{ia}^kM_k+f_{ia}^cP_c.
\eea
Consider a coset representative $\tilde G(z)=\tilde G(z_1,\dots z_p)$ of the coset space $G/H$. The Lie--algebra valued Maurer-Cartan one--forms
 \be\label{cartan}
\tilde G^{-1}d\tilde G=\omega^a P_a+w^i M_i,
\ee
are split into the forms $\omega^a$ belonging to the coset and $w^i$ which are associated with the subgroup $H$. The forms $\omega^a$ transform homogeneously under the group transformations. Due to this fact one can construct  a quadratic form on the coset space $G/H$, which holds invariant under the action of the group
\be\label{qua}
ds^2=K_{ab}\omega^a\omega^b.
\ee
The invariance of the metric under the group $G$ imposes the following restriction on the matrix $K_{ab}$ (see, e.g., Refs. \cite{ortin,cos})
\be\label{cond}
f^c_{i(a}K_{b)c}=0.
\ee

The geodesic equations on the coset space equipped with the metric (\ref{qua}) are derived from the action functional
\be\label{act}
S=\frac12 \int  K_{ab}\omega^a\omega^b d\lambda,
\ee
where it is assumed that in the Maurer-Cartan one--forms the differentials are replaced by velocities $\omega^a=\omega^a{}_\mu \dot{z}^{\mu}$, with $\mu=1,\dots,p$ and the dot denotes the derivative with respect to the parameter $\lambda$.
Taking into account the identity
\be\label{ide}
\delta \omega^a=\frac{d}{d\lambda}\delta X^a+\dot{z}^\nu\d z^\mu(\partial_\mu\omega^a{}_\nu-\partial_\nu\omega^a{}_\mu),
\ee
where $\delta X^a=\d z^\mu\omega^a{}_\mu$ and the Maurer-Cartan equations
\be\label{dual}
d\omega^a+\frac12f^a_{bc}\omega^b\wedge\omega^c+f^a_{ci}\omega^c\wedge w^i=0,
\ee
the variation of the action (\ref{act}) can be written in the following form
\be\label{var}
\d S=\int K_{ab}\omega^b\left(\frac{d}{d\lambda}\delta X^a-\frac12 f^a_{bc}\left(\d X^b\omega^c-\omega^b\delta X^c\right)-f^a_{ci}\left(\delta X^c w^i-\d X^i\omega^c\right)\right)d\lambda,
\ee
where it was also denoted $\d X^i=\d z^\mu w^i{}_\mu$. In the obtained expression the last term vanishes due to the condition  (\ref{cond}) on the matrix  $K_{ab}$. From the action variation (\ref{var})
one obtains the geodesic equations on the coset space $G/H$
\be\label{geodes}
K_{ab} \dot{\omega}^b=K_{bd}f^d_{ca}\omega^b\omega^c+K_{bd}f^d_{ia}\omega^b w^i.
\ee
Note that these equations are written in purely group--theoretic terms.

Focusing on the $l$--conformal Galilei algebra (\ref{algebra*}), Eq. (\ref{geodes}) lead to the geodesic equations associated with the metric (\ref{metr})
  \bea\label{ge}
&&
\frac{1}{2}\dot{\omega}_H=-\frac{1}{2}\omega_H\omega_D+(q-2l)S_{p,q+1}\omega^{(p)} \omega^{(q)},
\nonumber\\[2pt]
&&
\frac{1}{2}\dot{\omega}_K=\frac{1}{2}\omega_K\omega_D+qS_{p,q-1}\omega^{(p)} \omega^{(q)},
\nonumber\\[2pt]
&&
\dot{\omega}^{(p)}S_{p,n}=-(n-l)S_{p,n}\omega^{(p)}\omega_D-n S_{p,n-1}\omega^{(p)}\omega_H-(n-2l)S_{p,n+1}\omega^{(p)}\omega_K.
\eea
The equations above are invariant under the $l$-conformal Galilei group generated by \cite{cg}
\bea\label{transf}
&&
H=\frac{\partial}{\partial t},\qquad K=t^2\frac{\partial}{\partial t}+(1-2 t r)\frac{\partial}{\partial r}-2t(n-l)x^{(n)}_i \frac{\partial}{\partial x^{(n)}_i},
\nonumber\\[2pt]
&&
D=t\frac{\partial}{\partial t}-r\frac{\partial}{\partial r}-(n-l)x^{(n)}_i \frac{\partial}{\partial x^{(n)}_i},\qquad M_{ij}=x^{(n)}_i\frac{\partial}{\partial x^{(n)}_j}-x^{(n)}_j\frac{\partial}{\partial x^{(n)}_i},
\nonumber\\[2pt]
&&
C^{(m)}_i=B^{nm}\frac{\partial}{\partial x^{(n)}_i}, \quad B^{mn}=\sum_{s=0}^m \frac{{(-1)}^{n-s} m! (2l-s)!}{s! (m-s)! (n-s)! (2l-n)!} t^{m-s} r^{n-s},
\eea
where it is assumed that in the last formula the terms with $s>m$ and $s>n$ vanish. Note that
the Maurer--Cartan one--forms (\ref{MC}) hold invariant under the time translations and transformations, generated by vector generators $C^{(n)}_i$, while under the dilatations, special conformal transformations and rotations they transform as follows:
\bea
&&
\omega_H\rightarrow(1+2tb+c)\omega_H, \qquad \omega_K\rightarrow\left(1-2tb-c\right)\omega_K, \qquad \omega_D\rightarrow \omega_D-2b\omega_H,
\nonumber\\[2pt]
&&
\omega^{(n)}_i\rightarrow(\delta_{ij}-\delta_{ij}(n-l)(c+2bt)-\sigma_{ij}) \omega^{(n)}_j.
\eea

In order to construct the corresponding integrals of motion in the explicit form, we redefine the coordinates $x^{(n)}_i$
\be
x'^{(n)}_i=(B^{-1})^{np}x^{(p)}_i,
\ee
with the use of the matrix $(B^{-1})^{np}$ inverse to $B^{np}$
\be\label{B}
(B^{-1})^{np}=\sum_{q=n}^{2l} \frac{{(-1)}^{q-n} q! (2l-p)!}{n! (q-p)! (q-n)! (2l-q)!} t^{q-n} r^{q-p},
\ee
where it is assumed that the terms with $p>q$ and $n>q$ vanish.
In the new coordinate system $C^{(m)}_i$ generate shifts of the coordinate $x'^{(n)}_i$. It can be easily verified that such a coordinate system corresponds to using the coset of the form
\be\label{cos}
\tilde G=e^{i C^{(n)}_ix'^{(n)}_i} e^{i t H}e^{i r K}.
\ee
The metric associated with such a parametrization of the coset has the form (\ref{metr}) with
\be\label{omega'}
\omega^{(n)}_i= B^{np} dx'^{(p)}_i,
\ee
and $\omega_H$, $\omega_D$, $\omega_K$ unchanged. The equations of motion maintain their form (\ref{ge}).

The generators of the symmetry transformations in the new coordinate system read
\bea\label{sym}
&&
H=\frac{\partial}{\partial t}-(n+1)x'^{(n+1)}_i\frac{\partial}{\partial x'^{(n)}_i}, \qquad D=t\frac{\partial}{\partial t}-r\frac{\partial}{\partial r}-(n-l)x'^{(n)}_i\frac{\partial}{\partial x'^{(n)}_i},
\nonumber\\[4pt]
&&
K=t^2\frac{\partial}{\partial t}+(1-2tr)\frac{\partial}{\partial r}-(n-2l-1)x'^{(n-1)}_i\frac{\partial}{\partial x'^{(n)}_i}, \qquad C^{(n)}_i=\frac{\partial}{\partial x'^{(n)}_i}.
\eea

Now, we can write the integrals of motion in the explicit form
\bea\label{sym}
&&
H=2r^2\dot t+\dot r-(n+1)x'^{(n+1)}_iC^{(n)}_i, \qquad D=t(2r^2\dot t+\dot r)-r\dot t-(n-l)x'^{(n)}_iC^{(n)}_i,
\nonumber\\[4pt]
&&
K=t^2(2r^2\dot t+\dot r)+(1-2tr)\dot t-(n-2l-1)x'^{(n-1)}_iC^{(n)}_i, \quad C^{(n)}_i=B^{pn}S_{p,m}\omega_i^{(m)}.
\eea
When verifying that these quantities are conserved,  the following identities
\bea\label{id}
&&
\dot{B}^{mp}(B^{-1})^{pn}=(n-l)\omega_D\delta^{mn}+n\delta^{n-1,m}\omega_H+(n-2l)\omega_K\delta^{m,n+1},
\\[4pt]
&&
qB^{p,q-1}(B^{-1})^{qm}S_{pn}-r^2(n-2l)S_{m,n+1}-nS_{m,n-1}+m\leftrightarrow n=0,
\nonumber\\[4pt]
&&
(q-l)B^{pq}(B^{-1})^{qm}S_{pn}-(tr^2-r)(n-2l)S_{m,n+1}-tnS_{m,n-1}+m\leftrightarrow n=0,
\nonumber\\[4pt]
&&
(q-2l)B^{p,q+1}(B^{-1})^{qm}S_{p,n}-(1-t r)^2(n-2l)S_{m,n+1}-t^2nS_{m,n-1}+m\leftrightarrow n=0,
\nonumber
\eea
prove to be helpful. Above it is assumed that in the Maurer--Cartan one--forms the differentials are replaced by the velocities. It is straightforward to verify that (\ref{sym}) are the functionally independent integrals of motion for the second order system of equations (\ref{ge}). Note that, as compared to the geodesic equations associated with the Ricci--flat metrics in \cite{cg}, the geodesic equations (\ref{ge}) are defined on the coset space without introducing any extra variables. In this sense they represent the minimal model.

\vspace{0.5cm}

\noindent
{\bf 4. Einstein manifolds with the $l$-conformal Galilei isometry group}\\

\noindent
In this section we demonstrate that the coset space metric (\ref{metr}) can be extended to describe an Einstein manifold. Note that the first term in the quadratic form (\ref{metr}) represents the $AdS^2$ metric.
Following \cite{cg}, we shall rewrite the metric in the Poincar\'e coordinates which implies the redefinition of the temporal coordinate
\be
t\rightarrow \frac12 \left(t+\frac{1}{r}\right).
\ee
In this coordinate system the metric (\ref{metr}) reads
\bea\label{metric}
&&
ds^2=\a\left(r^2 d t {}^2-\frac{dr^2}{r^2}\right)+S_{n,m} \omega^{(n)}_i \omega^{(m)}_i,
\eea
where the forms $\omega^{(n)}_i$ are given in (\ref{MC}) with
\bea\label{ab}
&&
a^{(n)}_i=r(n-l) x^{(n)}_i -\frac 12 (n+1)  x^{(n+1)}_i-\frac 12 r^2 (n-2l-1)  x^{(n-1)}_i,
\nonumber\\[2pt]
&&
b^{(n)}_i=-\frac{1}{r} (n-l) x^{(n)}_i +\frac{1}{2r^2} (n+1)  x^{(n+1)}_i
-\frac 12 (n-2l-1)  x^{(n-1)}_i.
\eea
Above we multiplied the $AdS_2$ metric by a constant parameter $4\a$, which, as will be demonstrated below, is linked to the cosmological constant.

Following Ref. \cite{cg}, let us introduce an extra coordinate $y$ and extend the metric in the following way:
\be\label{metric1}
ds^2=\alpha\left(r^2 d t {}^2-\frac{dr^2}{r^2}\right)+S_{n,m}(y) \omega^{(n)}_i \omega^{(m)}_i+\epsilon dy^2,
\ee
where the matrix coefficients $S_{m,n}$ are now functions of $y$. Such a deformation of the metric does not break the invariance under the action of the $l$-conformal Galilei group.
Since one can always remove an arbitrary function of $y$ in front of the factor $dy^2$ by redefying $y$, only the arbitrariness in choice of the sign $\e=\pm1$ remains.

Now let us write down the explicit form of the components of the Ricci tensor
\be\label{xx_ricci}
R_{\{(m)i\}\{(n)j\}}=\frac{\e}{4} S^{p,q} \left(S'_{p,(m}S'_{n),q}-S'_{p,q} S'_{m,n}\right)\d_{ij}-\frac{\e}{2} S''_{m,n}\d_{ij}+\Omega_{mn}\d_{ij},
\ee
where the prime denotes the derivative with respect to $y$ and
\be\label{om}
\Omega_{mn}=\frac{1}{4\a} \left(q(p-2l){S}^{q,p}S_{q-1,m}S_{n,p+1}-m(n-2l)S_{m-1,n+1}\right)+(m\leftrightarrow n).
\ee
In order to simplify these equations, in what follows we assume that $S_{m,n}(y)=s(y) S_{m,n}$ with $s(y)$ to be determined and a constant matrix $S_{m,n}$. The requirement that
the Ricci tensor is proportional to the metric implies the vanishing of $\Omega_{mn}$ which yields the recurrence
\be\label{co}
S_{m,n}=\varepsilon_{mn} \frac{n}{m+1} S_{m+1,n-1},
\ee
with $\varepsilon_{mn}=\pm 1$. Similar to Ref. \cite{cg}, the algebraic conditions (\ref{co}) relate all the components of $S_{m,n}$ to a single independent parameter $S_{0,2l}$.

In view of the above, the components of the Ricci tensor take the form
\be\label{ric}
R_{mn}=\e\frac{s'^2}{s}\left(-\frac14 d (2l+1)+\frac12\right)S_{m,n}-\frac{\e}{2} s''S_{m,n},
\ee
where we denoted $R_{\{(m)i\}\{(n)j\}}=R_{mn}\d_{ij}$. The condition that the metric is Einstein, i.e. $R_{\m\n}+\lambda g_{\m\n}=0$, imposes the restriction on the function $s(y)$
 \be\label{eq}
s'^2-\frac{4\lambda \e}{d(2l+1)}s^2=0,
\ee
which follows from combining $yy$ and $xx$ components of the Einstein equations written above. The general solution of this equation has the form
\be\label{eq_s}
s(y)=C \exp \pm ky, \qquad k^2=\frac{4\lambda \e}{d(2l+1)},
\ee
with constant parameter $C$.
Due to the requirement for the metric to be real-valued, the product  $\lambda \e$ should be positive.

It is straightforward to verify that $R_{ty}$, $R_{ry}$, $R_{x y}$ vanish identically, while the remaining components read
\bea
&&
R_{tt}=R_{mn}a^{(m)}_i a^{(n)}_i+c r^2, \qquad R_{rr}=R_{mn}b^{(m)}_i b^{(n)}_i-\frac{c}{r^2},
\nonumber\\[2pt]
&&
R_{tr}=R_{mn}a^{(m)}_i b^{(n)}_i, \qquad R_{t\{(m) i\}}=R_{m p}a^{(p)}_i, \qquad R_{r\{(m) i\}}=R_{mp}b^{(p)}_i,
\eea
where
\be\label{const2}
c=\epsilon+\frac{ l(l+1)(2l+1) d \epsilon}{6}-\frac{d \epsilon}{4}{S}^{p+1,q-1} S_{p,q}  (p+1)(q-2l-1).
\ee
The metric (\ref{metric1}) thus obeys the equation $R_{\m\n}+\lambda g_{\m\n}=0$, if one relates the cosmological constant $\lambda$ and the parameter $\a$ entering to the metric as follows:
\be\label{a}
\a=-\frac{c}{\lambda}.
\ee
Similar to the Ricci--flat metrics constructed in \cite{cg},  the metric (\ref{metric}) is ultrahyperbolic. It describes a $[(2l+1)d+3]$-dimensional Einstein manifold, which involves one free parameter $C$ defined in Eq. (\ref{eq_s}).

\vspace{0.5cm}

\noindent
{\bf 5. Conclusion}\\

\noindent
To summarize, in this work we applied the general group theoretic construction so as to construct a novel dynamical realization of the $l$--conformal Galilei group in terms of geodesic equations on the coset spaces.
A peculiar feature of the geodesics is that all their integrals of motion, including the accelerations, are functionally independent. The dimension of coset grows with $l$ and to
each acceleration generator $C^{(n)}_i$ in the algebra there correspond extra spacetime dimensions.
We also extended the analysis in the recent work \cite{cg} and constructed
Einstein metrics with the $l$--conformal Galilei isometry group.

Due to the fact that the metrics constructed in this work involve the $AdS_2$ metric and have the $SO(2,1)$ symmetry subgroup, it is possible to extend the corresponding generators to those forming the Virasoro algebra, or more generally, the Virasoro--Kac--Moody type algebra. It would be interesting to investigate such models within the context of the nonrelativistic AdS/CFT--correspondence.

\vspace{0.5cm}

\noindent{\bf Acknowledgements}\\

\noindent
The author is grateful to A. Galajinsky for posing the problem, useful discussions, and reading the manuscript.
This work was supported by the MSE program Nauka under the project 3.825.2014/K and the RFBR grant 15-52-05022.

\end{document}